\documentclass[aps,pre,twocolumn,showpacs,superscriptaddress]{revtex4}
\usepackage{graphicx}
\usepackage{epsfig,xcolor,graphics}
\usepackage{amsmath,mathrsfs}
\usepackage{bm}
\usepackage{siunitx}
\usepackage{float}
\usepackage{tikz}
\usepackage{adjustbox}
\usepackage{pgfplots}
\pgfplotsset{compat=1.17}
\usepackage{comment}
\usepackage{hyperref}
\hypersetup{
    colorlinks   = true, 
    urlcolor     = blue, 
    linkcolor    = blue, 
    citecolor   = blue 
}

\graphicspath{ {./Figures/} }

\begin{document}
\title{Wet granular bed eroded by a dry granular flow}
\author{Lama Braysh}
\email{lama.braysh@umontpellier.fr}
\affiliation{LMGC, University of Montpellier, CNRS, France}
\author{Patrick Mutabaruka}
\email{patrick.mutabaruka@ifremer.fr}
\affiliation{Geo-Ocean, Univ. Brest, CNRS, IFREMER, Plouzané, France}
\author{Farhang Radjai}
\email{franck.radjai@umontpellier.fr}
\affiliation{LMGC, University of Montpellier, CNRS, France}
\author{Serge Mora}
\email{serge.mora@umontpellier.fr}
\affiliation{LMGC, University of Montpellier, CNRS, France}

\date{\today}

\begin{abstract}
Using three-dimensional Discrete Element Method (DEM) simulations, 
we investigate the erosion dynamics of a cohesive bed composed of wet 
spherical particles subjected to the shear flow of an overlying non-cohesive 
granular layer. Cohesion is modeled through a capillary attraction law, where 
the erosion process is governed by the irreversible rupture of liquid bridges 
at the interface. By systematically varying the liquid-vapor surface tension 
and the inclination angle of the bed, we analyze the influence of cohesive 
strength and flow intensity on the mass entrainment rate. Our results identify two 
distinct erosion regimes: a slow, stochastic regime driven by granular temperature 
fluctuations, and a fast, collective regime characterized by a global mechanical 
instability of the interface. We propose a robust scaling law for the surface erosion 
rate in the fast erosion regime, based on the interplay between two dimensionless 
parameters: the inertial number ($I$) and the cohesion index ($\xi$). 
This framework reveals that the threshold for the fast erosion regime is determined 
by the ratio of the geometric mean of the driving stresses (kinetic and pressure) to 
the cohesive resistance. These findings provide a comprehensive description of the 
coupling between inertial and capillary forces, offering a predictive tool for the stability 
of cohesive interfaces in sheared granular flows.


\end{abstract}

\maketitle

\section{Introduction} 
\label{sec:introduction}
The mechanical behavior of granular materials is significantly altered 
by the presence of cohesive forces. In unsaturated wet granular systems, 
for instance, the addition of a small amount of liquid induces cohesive 
interactions between the particles due to the surface tension of the liquid, 
enabling the formation of solid-like structures such as 
sandcastles \cite{herminghaus2005dynamics, mitarai2006wet}. These liquid bonds 
also generate lubrication forces due to the liquid viscosity \cite{vo2020evolution, Lefebvre2013Erosion}. 
From a rheological perspective, a homogeneous unsaturated wet granular material 
is characterized by its cohesive strength $\sigma_c$, which represents the maximum 
tensile stress the capillary bonds can sustain. The flow behavior and the resulting 
microstructure also depend on the confining pressure $\sigma_p$, which 
accounts for the external or lithostatic forces \cite{radjai2010force}, 
as well as the kinetic stress $\sigma_k$. This latter, defined as $\sigma_k = \rho d^2 \dot{\gamma}^2$, 
arises from the collective motion of particles and depends on the shear 
rate $\dot{\gamma}$, the particle density $\rho$, and the particle 
diameter $d$ \cite{bagnold1954experiments, da2005rheophysics}.

The rheology of homogeneous wet granular materials has been extensively 
studied both experimentally \cite{iveson2002dynamic, moller2007shear} and 
numerically \cite{rognon2006rheophysics, richefeu2006shear, Braysh2025JOR}. 
It is now well established that at least three dimensionless numbers govern the 
physics of these systems. The inertial number $I$ represents the interplay 
between particle inertia and confining stress, defined 
as \cite{da2005rheophysics, GDR-MiDi2004, Jop2006constitutive, kamrin2012nonlocal, berger2016scaling, khamseh2015flow, badetti2018rheology}
\begin{equation}
  I = \sqrt{\sigma_k / \sigma_p} \label{eq:def I}.
\end{equation}
Experimental and numerical studies show that the apparent friction coefficient $\mu$ 
and packing fraction $\Phi$ of steady flows of dry granular materials are unique 
functions of $I$ \cite{Jop2006constitutive, GDR-MiDi2004, pouliquen2006flow, forterre2008flows}. 
The other dimensionless parameters accounting for capillary cohesion and 
viscosity are the cohesion index \cite{vo2020evolution, berger2016scaling, khamseh2015flow, badetti2018shear}
\begin{equation}
  \xi = \sigma_c / \sigma_p \label{eq:def xi}
\end{equation}
and the Stokes number
\begin{equation}
St = \sigma_k / \sigma_v,
\end{equation}
where $\sigma_v$ relates to viscous stresses \cite{iveson2002dynamic, boyer2011unifying, trulsson2012transition}. 
Recent studies have shown that when inertial, cohesive, and viscous forces are 
simultaneously present, the system can be described by a combined visco-cohesive 
inertial number \cite{vo2020additive,vo2021erosion,amarsid2017viscoinertial,Tapia2022}.

The erosion of a cohesive granular material by a fluid or dry particle flow 
is of great importance in nature and industrial applications. A relevant example 
is the evolution of wet agglomerates embedded in a flow of non-cohesive granular 
materials \cite{hassanpour2007effect, hassanpour2007modeling, vo2020evolution}. 
In these contexts, an equilibrium is often reached between the cohesive forces 
maintaining the aggregate and the erosive action of non-cohesive particles. 
In industrial mixers or rotating drums, this motion is generated by the container's 
rotation, whereas in geophysics, it is driven by gravity. For instance, during 
a landslide, debris flows down a slope and erodes the underlying soil, which is 
often a cohesive granular medium \cite{mangeney2010erosion,He2024}. 
While the erosion of a granular bed by a liquid has been extensively modeled 
by transport models \cite{lajeunesse2010bed, charru2004erosion}, our understanding 
of erosion caused by a granular flow remains much more limited. 
In such cases, particles at the interface can be irreversibly detached wherever 
capillary bonds are broken \cite{vo2020evolution, Lefebvre2013Erosion, ennis1991microlevel, behjani2017investigation}.

\begin{figure}[!tbh]
  \includegraphics[width=0.6\columnwidth]{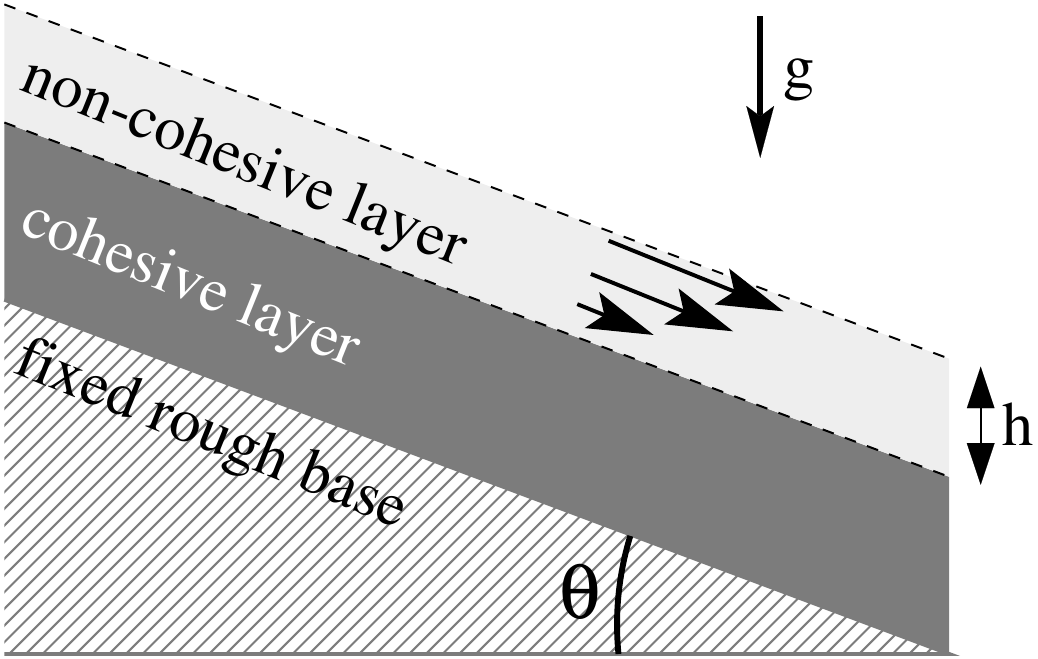}
  \caption{Schematic representation of the system, consisting of two granular layers 
  deposited on a rough, rigid plane inclined at an angle $\theta$ relative to the horizontal, 
  within a vertical gravitational field $g$. The upper layer comprises a non-cohesive 
  granular medium in a flow state driven by gravity. The underlying layer consists 
  of a cohesive granular medium, initially at rest. The erosion process occurs at the 
  interface between these two distinct phases, where particles are detached 
  from the cohesive bed and entrained into the overlying dry flow.}
\label{fig:general}
\end{figure}

In this work, we use simulations to investigate a bed of wet, cohesive particles overlaid by 
a shear flow of non-cohesive dry particles on an inclined plane (see Fig. \ref{fig:general}). 
When the inclination angle $\theta$ exceeds the avalanche angle of the dry 
layer but remains below the critical threshold required for the cohesive bed to flow, 
the system reaches a metastable state where the dry layer flows over a static 
cohesive substrate. The resulting erosion is a purely interfacial phenomenon, 
governed by the competition between the shear stress exerted by the 
dry flow and the stabilizing effects of cohesion $\sigma_c$ and confining pressure $\sigma_p$. 

As in the rheology of wet granular flows, the erosion dynamics at the interface 
between a cohesive bed or wet agglomerate 
and a flowing dry particles can in principle be described in terms of the 
dimensionless numbers $I$, $\xi$, and $St$.    
However, there are two fundamental differences between these  
two problems. First, the rheology of a cohesive granular material is a bulk behavior 
whereas erosion dynamics occurs at the interface between two media. Secondly, 
in rheology of cohesive flows the inertial and cohesive effects 
are simultaneously present everywhere inside the flow  
while for erosion the effects of attraction forces and viscous forces 
only play a role within the cohesive layer, controlling the cohesive 
strength against the shear flow, and the inertial effects 
concern only non-cohesive particles in the dry layer. 
For these reasons, we expect a nontrivial scaling behavior of the 
erosion rate with respect to system parameters.

In this article, we examine how the erosion process evolves as a function 
of key system parameters, including the dry layer thickness, the inclination angle, 
and the magnitude of the cohesive forces. We operate under conditions where 
the viscous stress ($\sigma_v$) remains negligible in comparison to the other 
governing stresses. This assumption allows us to isolate the competition between 
inertial and capillary forces without the complexity of lubrication effects. 
We demonstrate the existence of two distinct erosion regimes separated by a 
critical inclination angle $\theta_c$. For $\theta < \theta_c$, the erosion occurs 
on a grain-by-grain basis driven by rare, intermittent stochastic events; this is 
referred to as the ``slow erosion regime''. Conversely, for $\theta > \theta_c$, the 
system enters a ``fast erosion regime'', triggered when the static equilibrium conditions 
of the cohesive bed are no longer satisfied, leading to a massive entrainment of particles.

The article is organized as follows. In Section \ref{sec:theory}, we detail 
the physical processes leading to the two aforementioned erosion regimes. 
This theoretical framework allows us to derive a scaling law for the critical angle 
$\theta_c$ and to identify the characteristic timescales involved in the fast erosion regime. 
In Section \ref{sec:NumericalProcedures}, we present the numerical method based 
on Discrete Element Method (DEM) simulations used to investigate the erosion 
of a cohesive bed by a non-cohesive flow on an inclined plane. 
Section \ref{sec:simulations} presents the simulation results, focusing specifically 
on the fast erosion regime. We establish a scaling law for the surface erosion rate, 
providing a unified framework for predicting mass entrainment in cohesive granular 
systems within this regime. Finally, Section \ref{sec:conclusion} provides a concluding 
discussion and outlines future perspectives for this work.

\section{Theoretical Framework}\label{sec:theory}

In this section, we develop a theoretical analysis of the erosion process to 
derive the governing laws for the fractional entrainment rate ($K_e$). We 
distinguish between the stochastic mechanisms prevalent at low stresses 
and the collective mechanical instability occurring at higher shear rates. 
These analytical derivations provide the scaling foundations that will be 
compared against the results of DEM simulations presented in Section \ref{sec:simulations}.

\subsection{Stability Analysis}

In the mechanical description of granular systems, the resistance to 
deformation is characterized by the ratio of the shear stress $\sigma_\tau$ to the 
normal stress $\sigma_n$. 
The static friction coefficient $\mu_s$ represents the maximum stress ratio 
that a granular medium can sustain before yielding. The criterion for stability 
is given by $\sigma_\tau \leq \mu_s \sigma_n$,  where $\mu_s$ defines the 
threshold required to initiate flow from a state of rest. 
The dynamic friction coefficient $\mu$ characterizes the medium once it is 
in a state of flow. In this case, the stress ratio $\sigma_\tau /\sigma_n$ is no longer 
a constant threshold but depends on the flow conditions (notably the shear rate).

While $\mu_s$ defines the stability of dry grains, the presence of inter-particle 
bonds in a cohesive bed enhances this resistance. Following recent developments 
in wet granular flows \cite{vo2020additive}, the static threshold $\mu_{c}$ of the 
cohesive bed can be expressed as:
\begin{equation}
  \mu_{c} = \mu_s(1 + a \xi), 
  \label{eq:mu_c}
\end{equation}
where $a$ is a material-dependent constant and $\xi$ is the cohesion index. 
In granular rheology, a  distinction exists between $\mu_s$ and the 
quasi-static limit $\mu_0$ (the limit of the effective friction as the flow 
velocity tends toward zero). While $\mu_s$ is typically slightly larger than 
$\mu_0$ due to grain interlocking at rest, for the sake of simplicity and as 
a first-order approximation, we hereafter assume $\mu_s \simeq \mu_0$.

The cohesive strength $\sigma_c$ within the bed arises from capillary bridge forces:
\begin{equation}
  \sigma_c \sim \frac{2 \pi \gamma_s \cos(\Theta_\ell)}{d},
\end{equation}
where $\gamma_s$ is the surface tension and $\Theta_\ell$ is the liquid-solid 
contact angle. By normalizing this strength by the confining pressure $\sigma_p$ 
exerted by the overlying dry layer, we define the cohesion index $\xi$:
\begin{equation}
\xi = \frac{\sigma_c}{\sigma_p} = \frac{2 \pi \gamma_s \cos(\Theta_\ell)}{\rho \phi g h d \cos\theta}, 
\label{eq:xi}
\end{equation}
where $h$ is the height of the non-cohesive medium, $\rho$ the 
particle density, and $\phi$ the solid volume fraction.

The non-cohesive layer is assumed to be in a flow state. At the interface, 
the ratio of shear stress to normal stress is governed by the inertial 
number $I$, defined as $I = \sqrt{\sigma_k / \sigma_p}$. 
The effective friction $\mu(I)$ of this layer is described by the $\mu(I)$ rheology \cite{Jop2006constitutive}:
\begin{align}
  \mu(I) &= \mu_0 + \frac{\mu_1 - \mu_0}{1+I_0/I} \\
  &\simeq \mu_0(1+\alpha I), 
  \label{eq:mu(I)}
\end{align}
where
\begin{equation}
  \alpha \simeq (\mu_1 - \mu_0)/(\mu_0 I_0).
  \label{eqn:def alpha}
\end{equation}
  While for a steady-state flow on an inclined plane $\mu(I) = \tan\theta$, 
  we also consider transient regimes where $I$ must be calculated from the 
  local shear rate $\dot{\gamma}$ at the interface:
\begin{equation}
I = \sqrt{\frac{d^2 \dot{\gamma}^2}{\phi g h \cos\theta}}. 
\label{eq:I}
\end{equation}

The transition from the slow erosion regime to the fast erosion regime 
occurs when the macroscopic stability limit of the cohesive bed is reached. 
This boundary is defined by the condition where the stress ratio exerted 
by the dry flow matches the cohesive yield threshold:
\begin{equation}
\mu(I) = \mu_{c}.
\end{equation}
Substituting the linearized form of $\mu(I)$ from Eq. \eqref{eq:mu(I)} and 
the definition of $\mu_{c}$ from Eq. \eqref{eq:mu_c}, we obtain:
\begin{equation}
\mu_0(1 + \alpha I) = \mu_0(1 + a \xi), 
\end{equation}
which simplifies to a critical ratio between the inertial number and the cohesion index:
\begin{equation}
\frac{I}{\xi} = \frac{a}{\alpha}.
\label{eqn:Ixi}
\end{equation}
This stability criterion leads to two distinct physical scenarios:
\begin{itemize}
    \item Slow Erosion ($I/\xi < a/\alpha$): The bed is macroscopically stable. 
    Erosion occurs via stochastic, grain-by-grain detachment driven by velocity fluctuations within the dry flow.
    \item Fast Erosion ($I/\xi > a/\alpha$): The mean shear stress 
    exceeds the bed's resistance, leading to a progressive failure of the cohesive layers.
\end{itemize}

By combining Eqs. \eqref{eq:I} and \eqref{eq:xi}, the condition 
Eq. \eqref{eqn:Ixi} can be expressed as a critical inclination angle, 
provided the local shear rate $\dot{\gamma}$ is known. In the following sections, 
$\dot{\gamma}$ is determined from the velocity profiles within the non-cohesive layer.

\subsection{Slow erosion regime} 
\label{sec:slow}

In the slow erosion regime, the stress ratio $\sigma_\tau/\sigma_p$ applied 
by the non-cohesive flow at the interface remains below the cohesive yield 
threshold $\mu_{c}$ defined in Eq. \eqref{eq:mu_c}. Consequently, the cohesive 
bed is macroscopically stable and does not undergo global failure. 
However, at the particle scale, the interface is subjected to a continuous 
bombardment by the overlying dry grains. This bombardment can trigger 
individual detachment events through a stochastic process governed by 
the competition between the local bond energy $E_b$ and the fluctuating 
kinetic energy $e_k$ of the dry particles.

The energy required to rupture a cohesive bond can be scaled as 
$E_b \sim \sigma_c d^3$. Conversely, the mean kinetic energy of a grain in the dry 
flow, $\langle e_k \rangle$, is proportional to the kinetic stress $\sigma_k$:
\begin{equation}
    \langle e_k \rangle \sim \sigma_k d^3 \sim \rho d^5 \dot{\gamma}^2, 
\end{equation}
In this regime, since the mean shear stress is insufficient to break the bed, 
the mean kinetic energy is typically lower than the bond energy ($\langle e_k \rangle < E_b$).


Under this sub-threshold condition, an erosion event can occur through
kinetic energy fluctuations:  the instantaneous velocity $\mathbf{u}$ of 
a dry grain at the interface fluctuates due to collisions. In dense granular 
flows, the distribution of these fluctuations often exhibits high-energy tails. 
A detachment is triggered if an impacting grain belongs to this tail, 
such that its instantaneous kinetic energy $e_k = \frac{1}{2}m u^2$ exceeds $E_b$.

The probability of detachment $P_{det}$ depends on the probability 
density function of the energy fluctuations $P(e_k)$. Assuming an exponential 
decay for the high-energy tails, as frequently observed in sheared granular media, 
the detachment probability can be modeled by an Arrhenius-like activation law:
\begin{equation}
    P_{det} \propto \exp\left( - \beta \frac{E_b}{\langle e_k \rangle} \right) = \exp\left( - \beta \frac{\sigma_c}{\sigma_k} \right),
\end{equation}
where $\beta$ is a non-dimensional coefficient related to the flow 
microstructure and the efficiency of energy transfer. 
Hence, as long as the dry layer is in motion, $P_{det}$ remains non-zero.

We define the surface erosion rate $K_e$ as the interfacial detachment frequency:
\begin{equation}
  K_e = \frac{1}{N_s} \frac{dN_e}{dt}
  \label{eq:Ke}
\end{equation}
where $N_s$ is the number of particles currently exposed at the 
wet-dry interface. $K_e$ can be estimated by the frequency of interface 
attempts $f \sim \dot{\gamma}$ over the number of particle sites  multiplied by the detachment probability:
    \begin{equation}
        K_e \sim \dot{\gamma}\exp\left( - \beta \frac{2\pi \gamma_s \cos\Theta_\ell}{\rho d^3 \dot{\gamma}^2} \right). 
        \label{eq:scaling_slow}
    \end{equation}
In summary, the slow regime is characterized by a frequency-dependent, 
thermally-activated-like process where the erosion rate is extremely 
sensitive to the ratio $\sigma_c / \sigma_k$.



\subsection{Fast erosion regime} 
\label{sec:fast}

When the stress ratio at the interface exceeds the cohesive yield threshold 
($I/\xi > a/\alpha$), the system enters the fast erosion regime. Unlike the 
stochastic process described in the previous section, this regime is characterized 
by a global mechanical instability of the interface. In this state, the mean 
shear stress exerted by the flowing dry layer is sufficient to overcome 
the macroscopic resistance of the cohesive bed. Consequently, erosion is no longer 
limited to rare, intermittent events but involves in addition a continuous and 
correlated failure of the cohesive layers.

In the fast regime, the mean-field driving forces (pressure $\sigma_p$ 
and shear stress $\sigma_\tau$) dominate the local bonding forces. 
While statistical fluctuations in the particle velocities still exist, they 
no longer play a leading role because the \textit{average} energy transfer 
from the dry flow to the bed is sufficient to trigger detachment. 
The process remains fundamentally interfacial, but it manifests itself as a 
progressive destabilization where the shear from the non-cohesive flow 
penetrates the cohesive bed, effectively entraining particles in a collective manner.

A defining feature of this regime is its characteristic timescale, $\tau_f$, 
which is expected to be significantly shorter than the long waiting 
times associated with the activated detachment in the slow regime. 
To estimate $\tau_f$, we consider the dynamics of a non cohesive particle 
of mass $m \sim \rho d^3$ located at the interface. Under the action of 
the mean confining pressure $\sigma_p$, the particle is subjected to a 
characteristic force $F_p \sim \sigma_p d^2$. We define $\tau_f$ as the time 
required for this force to accelerate the particle until it acquires a kinetic 
energy equal to the cohesive bond energy, $E_b \sim \sigma_c d^3$.
Following the second law of motion, we have $F_p \cdot \tau_f \sim m \cdot v$, 
where $v$ is the velocity such that $\frac{1}{2}m v^2 \sim \sigma_c d^3$. 
This yields $v \sim \sqrt{\sigma_c/\rho}$. Substituting this into the momentum balance, we obtain
\begin{equation}
    \tau_f \sim \frac{m v}{F_p} \sim \frac{(\rho d^3) \sqrt{\sigma_c/\rho}}{\sigma_p d^2} = 
    \frac{d \sqrt{\rho \sigma_c}}{\sigma_p}.
\end{equation}
By introducing the dimensionless numbers $I = \dot{\gamma} d / \sqrt{\sigma_p/\rho}$ 
and $\xi = \sigma_c/\sigma_p$, this timescale can be rewritten in terms of the macroscopic flow parameters:
\begin{equation}
    \tau_f \sim \frac{I \sqrt{\xi}}{\dot{\gamma}}.
\end{equation}
The derivation of $\tau_f$ suggests that the surface erosion rate 
in this regime, $K_e$ (Eq. \eqref{eq:Ke}), scales with the inverse of this 
characteristic time. Since the onset of this regime is governed by the ratio $I/\xi$, 
we expect the erosion rate to be a function of this parameter:
\begin{equation}
  K_e \propto \tau_f \mathcal{F}\left(\frac{I}{\xi}\right) 
  = \frac{\dot{\gamma}}{d^2 I \sqrt{\xi}} \mathcal{F}\left(\frac{I}{\xi}\right),
  \label{eqn:scaling fast}
\end{equation}
where $\mathcal{F}$ is a scaling function to be determined. 

While this mean-field analysis provides the governing scales 
of the problem, the complex spatial and temporal evolution of the interface 
during mass entrainment—particularly the feedback between the increasing 
thickness of the dry layer and the erosion of the bed—requires a more detailed 
investigation. In the following sections, we explore the functional form of $\mathcal{F}$ from 
DEM simulations to characterize the evolution of the erosion rate across 
a wide range of system parameters.

\section{Numerical procedures}
\label{sec:NumericalProcedures}

\subsection{Simulation method}

Numerical simulations are performed using an in-house software, 
cFGd3D \cite{Mutabaruka2014Initiation, Mutabaruka2019Effects}, which is based 
on the DEM   
\cite{cundall1979discrete, herrmann1998modeling, thornton2000quasi, radjai2011discrete}. 
This method falls under the broader category of particle dynamics 
techniques, similar to molecular dynamics  used for simulating 
molecular systems \cite{allen1987computer} and contact dynamics (CD) 
used to model systems with frictional contacts \cite{moreau1994sorne, radjai2009contact, radjai2018multi}.

In the basic DEM framework, particles interact via local force 
laws \cite{herrmann2013physics}, with these forces depending on contact strains, 
which are determined in terms of the relative displacements of the particles. 
The code implements a velocity-Verlet scheme \cite{allen1987computer} 
for step-wise integration of Newton's second law equations of 
motion for spherical particles.

The total interaction force $\bm{f}$ between two particles consists 
of normal and tangential components $f_n$ and $\bm{f}_t$, respectively:
\begin{equation}
\bm{f} = f_n \bm{n} + \bm{f}_t,
\end{equation}
where $\bm{n}$ is the contact normal. The normal force $f_n$ is the 
sum of the visco-elastic force $f_{ed}$ and the capillary force $f_{cap}$:
\begin{equation}
f_n = f_{ed} + f_{cap}.
\end{equation}
The distance $\delta_n$ represents elastic deflection (negative for overlap) 
or capillary bond length (positive for gap). The visco-elastic 
force $f_{ed}$ is the sum of linear elastic repulsion $f_n^e = - k_n \delta_n$ 
and viscous damping $f_n^d = - \gamma_n \dot{\delta}_n$, 
where $\gamma_n$ is the damping coefficient. 
The total normal force $f_{ed}$ is given by:
\begin{equation}
f_{ed} = 
\begin{cases} 
- k_n \delta_n - \gamma_n \dot{\delta}_n  & \text{if } f_{ed} > 0 \text{ and } \delta_n \leq 0, \\
0 & \text{otherwise}.
\end{cases}
\end{equation}
This ensures that $f_{ed}$ remains positive for dry contacts, 
but may become negative (attractive) when capillary forces are considered.

The liquid is assumed to be in the pendular state, where small liquid 
bridges join particles. The capillary stress increases with the amount 
of liquid in this state \cite{richefeu2006shear}, while it remains nearly 
constant in the funicular state \cite{scheel2008liquid, Delenne2015}. 
We model the capillary interaction as a binary force law, 
which provides a good approximation even at higher liquid volumes.

Capillary bonds are persistent and reversible unless the gap distance exceeds a rupture distance 
$d_{rupt}$, beyond which the liquid bridge collapses.
The rupture distance depends on the bond volume $V_b$ and is given by:
\begin{equation}
d_{rupt} = \left( 1+ \frac{\theta_{\ell}}{2} \right) V_b^{1/3},
\end{equation}
where $\theta_{\ell}$ is the particle-liquid-gas contact angle. 
This assumption simplifies DEM simulations, avoiding complex thermodynamic 
or hydrodynamic considerations \cite{Braysh2024JFM,Braysh2025JOR}.

The capillary force $f_{cap}$ is derived from the Young-Laplace 
equation \cite{Mikami1998}, approximated as:
\begin{equation}
f_{cap} = 
\begin{cases}
- \pi d \gamma_s \cos \theta_{\ell} & \text{for } \delta_n < 0, \\
- \pi d \gamma_s \cos \theta_{\ell} e^{-\delta_n/\lambda} & \text{for } 0 \leq \delta_n \leq d_{rupt},
\end{cases}
\end{equation}
where $\lambda \simeq 0.9 \left( \frac{2V_b}{d} \right)^{1/2}$.
 
For tangential forces, we use a linear elastic law with Coulomb friction:
\begin{equation}
\bm{f}_t = 
\begin{cases} 
- k_t \bm{\delta}_t - \gamma_t \dot{\bm{\delta}}_t & \text{if } || \bm{f}_t || \leq \mu_p f_n^e, \\
- \mu_p f_n^e \frac{\dot{\bm{\delta}}_t}{|| \dot{\bm{\delta}}_t ||} & \text{otherwise},
\end{cases}
\end{equation}
where $\mu_p$ is the friction coefficient, $k_t$ is the 
tangential stiffness, and $\gamma_t$ is the tangential damping 
parameter. $\bm{\delta}_t$ is the cumulative tangential displacement, 
and $\dot{\bm{\delta}}_t$ is the relative tangential velocity.

\begin{table}[!tbh]
\caption{Simulation parameters}
\begin{tabular}{lc}
Description						&	Value	\\
\hline
Number of particles, $N_p$ 			&	8,000	\\
Particle diameter, $d$				&	\SI{3e-4}{\meter}\\
Particle density, $\rho$				&	\SI{2.5}{\gram \per \cubic \centi \meter}	\\
Normal stiffness, $k_n$				&	\SI{1e2}{\newton \per \meter}	\\
Normal damping, $\gamma_n$			&	\SI{3e-3}{\newton \second \per \meter}	\\
Tangential stiffness, $k_t$				&	\SI{8e1}{\newton \per \meter}	\\
Tangential damping, $\gamma_t$		&	\SI{3e-3}{\newton \second \per \meter}	\\
Friction coefficient, $\mu_p$				&	{0.5}\\
Contact angle, $\Theta_\ell$				&	\SI{5}{\degree}	\\
Time step, $dt$					&	\SI{1e-6}{\second} \\
De-bonding distance, $d_{rupt}$			&       \SI{3e-5}{\meter} \\
Inclination angle, $\theta$				&	$\left[\SI{25}{\degree} - \SI{32}{\degree}\right]$	\\
Liquid surface tension, $\gamma_s$		&	$\left[\SI{8e-3}{\newton \per \meter} - \SI{4e-2}{\newton \per \meter}\right]$ \\

\end{tabular}
\label{Tab:ConstSimParam}
\end{table}

\subsection{System preparation}
\label{sec:SamplePreparation}

The granular bed is composed of $N_p=8,000$ non-cohesive monodisperse 
spherical particles with a diameter $d = 3 \times 10^{-4}$ m and 
a density $\rho = \SI{2500}{\kilo \gram \per \cubic \meter}$. 
The particles are dropped into a rectangular domain under the 
effect of gravity. 
The domain has dimensions of length $L_x = 26d$, 
width $L_y = 10d$, and height $L_z = 28d$. Once the particles reach 
their equilibrium state, the lateral walls are removed and we apply 
periodic boundary conditions along the $x$ and $y$ directions. 

\begin{figure}[!tbh]
\includegraphics[width=0.9\columnwidth]{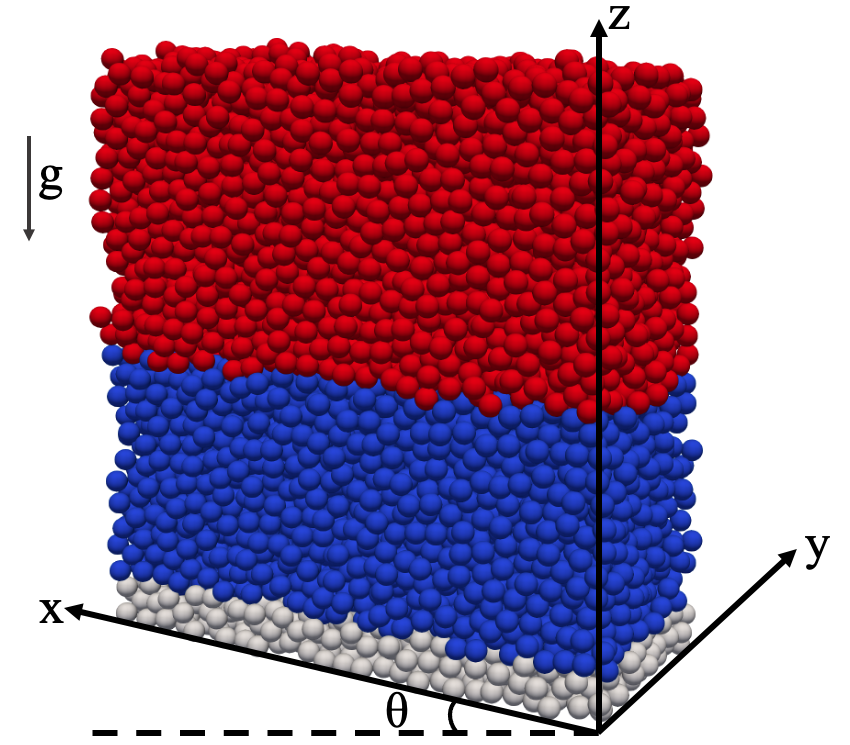}
\caption{A snapshot of the granular bed in its initial configuration, 
consisting of 3,500 wet particles (blue) overlaid by a non-cohesive 
bed of 4,000 dry particles (red). Additionally, 1000 immobile particles are 
stuck to the bottom of the sample (grey). 
The gravity vector is rotated to an angle $\theta$ (above the angle of repose)  
to trigger the flow of the particles in the non-cohesive layer.} 
\label{fig:SampleImage}
\end{figure}

The simulated system is divided into three layers, as illustrated in Fig. \ref {fig:SampleImage}: 
a rough fixed base at the bottom of the sample, consisting of densely packed particles 
with a height nearly $3d$ to prevent sliding (grey layer), a cohesive layer  
of thickness $12d$ (3,500 wet particles in red), and non-cohesive 
layer of thickness $13d$ (4000 dry particles in blue).  
The cohesive forces are activated between particles that belong to the cohesive layer and are 
separated by a distance less than or equal to the debonding distance $d_{rupt}=0.1d$. 
This ensures a uniform distribution of the wetting fluid on all particles. 

In this work, we assume irreversible breakage of capillary bonds to ensure 
that the eroded layer is not regenerated again. Indeed, in a large system 
of particles, once a particle gets eroded away from the wet layer,  the liquid it carries is rapidly 
dispersed by the flow so that the particle becomes dry. This assumption implies an irreversible 
extraction of particles from the wet layer. In the simulations, the status of a particle which has lost all its 
capillary bonds is changed from ``wet" to ``dry". The debonding distance in all 
simulations is set to $d_{rupt}=0.1d$. Note also that the interactions between 
wet and dry particles are assumed to be noncohesive. In other words, the 
capillary bonds are effective only between wet particles. 

The system is tilted by rotating the gravity vector of coordinates ($g \sin(\theta), 0, g \cos(\theta))$. 
We vary the inclination angle $\theta$ and  
liquid surface tension $\gamma_s$ in the cohesive layer. 
The ranges of values of these parameters and all constant system parameters  
used in this study are summarized  in Table~\ref{Tab:ConstSimParam}.

\section{Simulation results}
\label{sec:simulations}

\subsection{Phenomenology}

\begin{figure}[!tbh]
\begin{center}
\includegraphics[width=1.0\columnwidth]{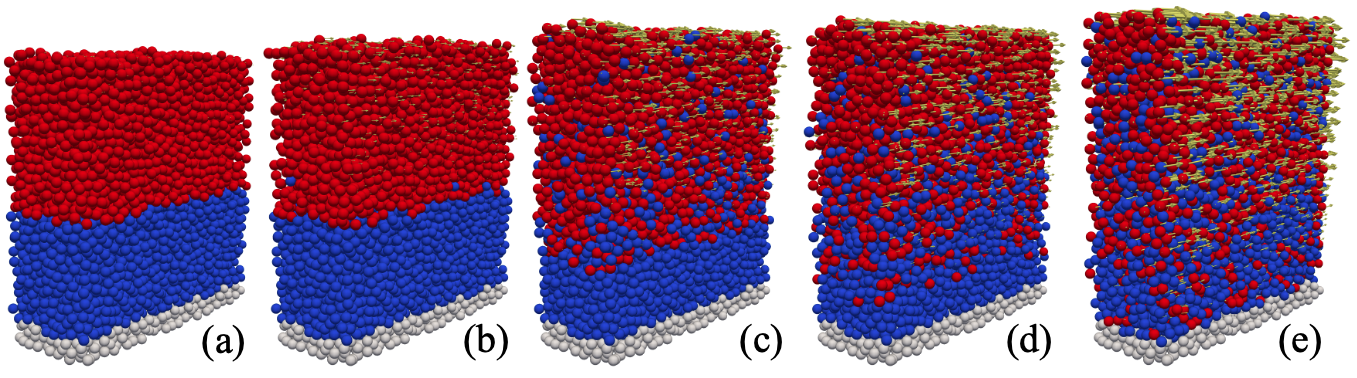}
\end{center}
\caption{Snapshots showing the time evolution of the erosion process for an 
inclination angle \(\theta = \SI{32}{\degree}\). Velocity vector thickness is proportional to particle velocity. 
Time snapshots: (a) \(t = 0.1\) s, (b) \(t = 0.5\) s, (c) \(t = 8\) s, (d) \(t = 10\) s, and (e) \(t = 13\) s. 
The corresponding velocity profiles are shown in Fig. \ref{fig:VelocityProfiles}.
In these snapshots, the particles keep their initial color as shown in Fig. \ref{fig:SampleImage} 
although the blue eroded particles (dispersed inside the dry layer) are no more wet. }
\label{fig:ErosionImage}
\end{figure}

The erosion  process is a consequence of the irreversible rupture of 
capillary bonds at the interface between the wet and dry layers. 
This is illustrated in Fig. \ref{fig:ErosionImage}, 
providing a typical example of the system's evolution from its 
initial tilting at time $ t = 0 $ to the completion of the erosion. 
The cohesive bed of initially wet particles, indicated by blue particles in 
Fig. \ref{fig:ErosionImage}, is placed at the bottom, just above a rough, 
immobile base represented by gray grains. 
In these snapshots the particles keep their initial color.  
For this reason, the blue eroded particles, which are 
dispersed inside the dry layer,  are no more wet.

The inclination angle $\theta$ is chosen 
to be larger than the avalanche angle $\theta^d_0$ (with $\tan \theta^d_0 = \mu_0$) 
of the non-cohesive dry medium but 
smaller than the avalanche angle $\theta^w_0$ (with $\tan \theta^w_0 = \mu_c$) of the wet cohesive layer. 
As a result, the cohesive bed does not flow while the non-cohesive 
layer flows over it. 
The interface between the two layers is the locus of erosion where 
wet particles are gradually detached and carried aways by the dry flow. 
Consequently, the cohesive phase thins over time while  
eroded particles spread evenly into the non-cohesive 
region, causing the thickness of the non-cohesive layer to increase. 
This behavior is observed across a broad range of the values of the liquid surface 
tension and tilt angle.

\begin{figure}[tbh]
\begin{center}
\includegraphics[width=0.9\columnwidth]{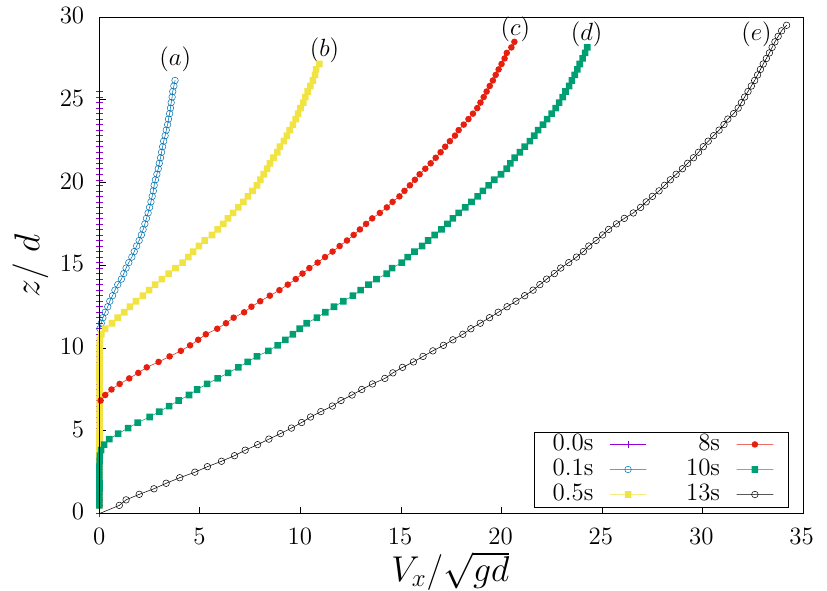}
\end{center}
\caption{
Normalized velocity profiles, using the characteristic velocity \(\sqrt{gd}\), 
along the direction of flow as a function of normalized height \(z/d\) for a system of 
surface tension \(\gamma_s = \SI{0.03}{\newton \per \meter}\) and 
inclination angle \(\theta = \SI{32}{\degree}\) at several instants of time corresponding 
to the snapshots (a-e) in Fig. \ref{fig:ErosionImage}. 
The flow is initially at rest (\(t = 0\) s), with the dry system beginning to flow while 
the velocity in the cohesive bed remains zero. As erosion progresses, the thickness 
of the static wet layer decreases. In the final stage, the entire sample flows 
as the cohesive layer becomes fully eroded.} 
\label{fig:VelocityProfiles}
\end{figure}

Fig. \ref{fig:VelocityProfiles} shows velocity profiles at 
several instants during the erosion of a wet bed. These profiles are 
calculated by averaging the particle velocities within layers parallel 
to the slope, each with a thickness of $\delta z = d$, 
where $d$ is the particle diameter. The average velocity is calculated over 
a time interval $\delta t$ that is much shorter than the system's characteristic 
evolution time. Only the $x$-component of the velocity (along the inclined plane) 
has a non-zero average, as shown in Fig. \ref{fig:VelocityProfiles}. Initially, 
the velocity profile is zero, as the system starts at rest. As the erosion proceeds, 
the velocity profile splits into two distinct zones. The first zone, at the bottom 
of the sample, corresponds to the cohesive bed where the velocity is zero. 
Note that particle velocities within the rough 
base are fixed at zero and are not shown ($z < 0$). Above a certain height, 
the average particle velocity becomes non-zero. This height marks the separation 
interface between the cohesive and non-cohesive parts, which evolves over 
time. Initially, this interface is at the midpoint of the system but gradually 
moves downward as the erosion process continues, eventually reaching 
zero when the process is complete.

\begin{figure}[!tbh]
\includegraphics[width=0.9\columnwidth]{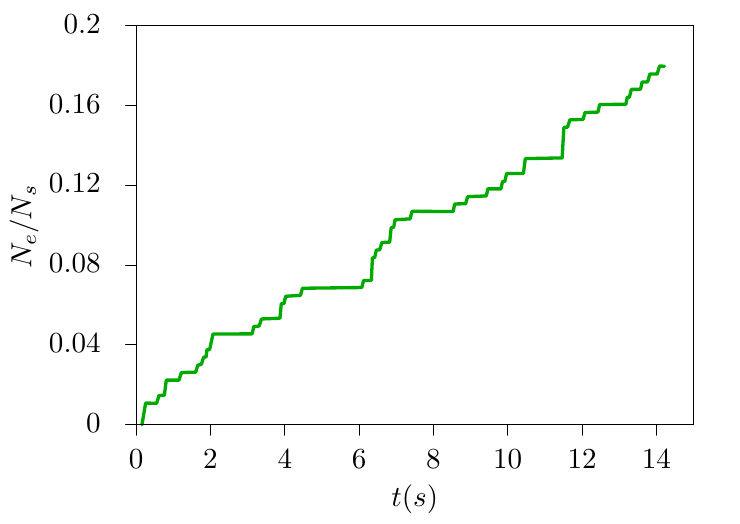}
\caption{Temporal evolution of the number of eroded particles $N_e$, 
normalized by the average number of particles $N_s$ located at the 
interface between the cohesive and non-cohesive zones. 
The data are obtained from DEM simulations with a surface tension 
$\gamma_s = 0.03$~N/m and an inclination angle $\theta = 25^\circ$.}
\label{fig : slow_regime}
\end{figure}

\begin{figure}[!tbh]
\includegraphics[width=0.9\columnwidth]{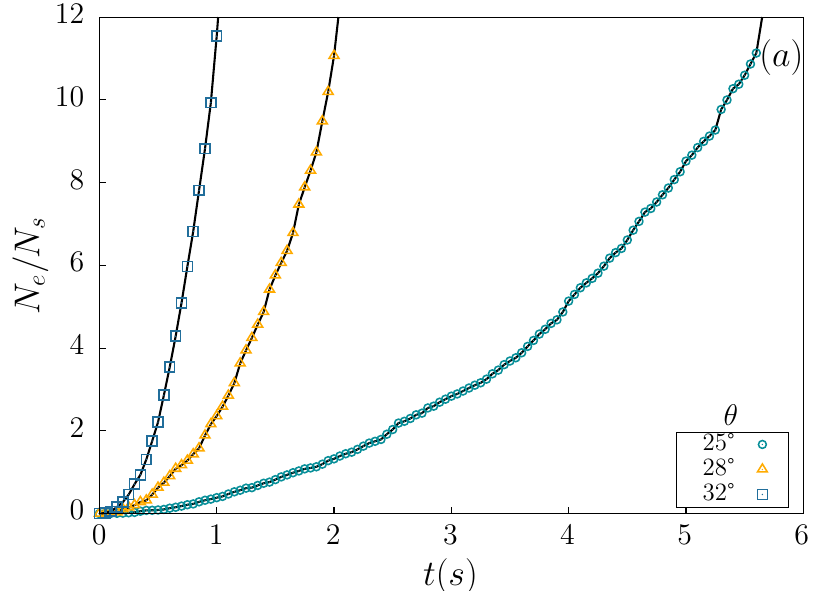}
\includegraphics[width=0.9\columnwidth]{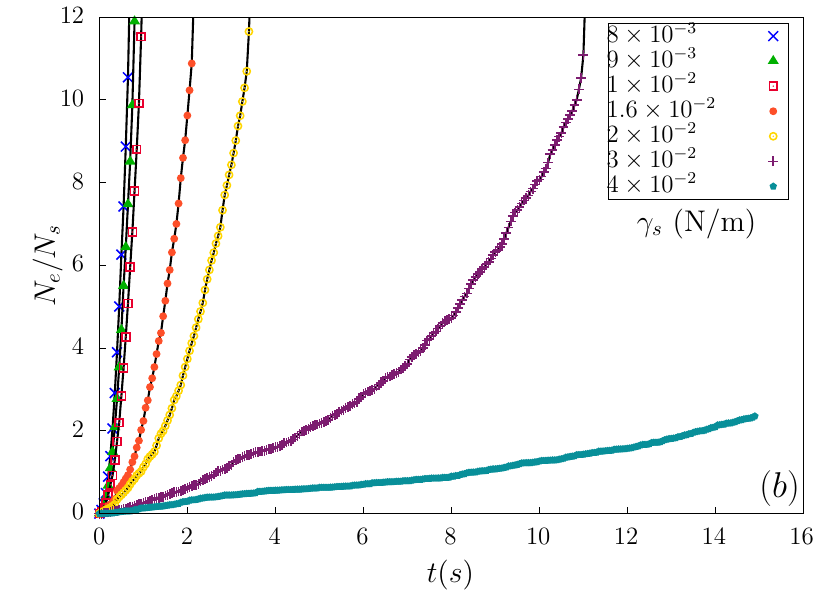}
\caption{Time evolution of the number $N_e$ of eroded particles normalized by 
the average number $N_s$ of particles at the interface between the cohesive and non-cohesive 
zones. (a) For three values of the inclination angle $\theta$ 
with $\gamma=\SI{0.01}{\newton \per \meter}$, and (b) different values of the 
liquid surface tension with $\theta=\SI{32}{\degree}$.
} 
\label{fig : time evolution}
\end{figure}

As an illustration, Figs.  \ref{fig : time evolution} and \ref{fig : slow_regime}  display the evolution of the 
total number $ N_e$ of eroded particles as a function of time for various 
values of the surface tension and inclination angle.
Since the erosion process takes place at the interface between the cohesive 
and non-cohesive regions, we normalize $ N_e $ by  the average 
number $ N_s $ of particles at the interface. The latter remains constant during the process.  
Significant differences are observed between the results presented 
in Fig. \ref{fig : slow_regime} and those in Fig. \ref{fig : time evolution}. 
Fig. \ref{fig : slow_regime} characterizes the slow erosion regime, whereas 
the curves in Fig. \ref{fig : time evolution} correspond to the fast erosion 
regime, as it will be discussed in detail in Section \ref{sec:sim fast}. 
For a given erosion duration, the values of the normalized number of eroded 
particles, $N_e/N_s$, are several orders of magnitude lower in Fig. \ref{fig : slow_regime} 
than in Fig. \ref{fig : time evolution}. The discrete ``staircase" jumps visible in Fig. \ref{fig : slow_regime} 
correspond, at their smallest scale, to the detachment of single particles. 
In some instances, the jump amplitude is an integer multiple of this 
fundamental unit, suggesting that the removal of a single grain can locally 
destabilize neighboring particles.

In the present study, we do not focus extensively on the slow erosion regime. 
Simulating this regime poses a significant numerical challenge due to 
the vast separation of timescales between the fast particle dynamics 
within the non-cohesive flow and the extremely low frequency of erosion 
events. A systematic investigation of the slow erosion regime will be the subject of a future publication. 
Nevertheless, the data in Fig. \ref{fig : slow_regime} allow for a preliminary 
estimation of the dimensionless coefficient $\beta$ introduced in Eq. \eqref{eq:scaling_slow}. 
Analysis of the velocity profiles within the dry flow provides the local 
shear rate at the interface, $\dot{\gamma} \approx 70$~s$^{-1}$ for 
the case shown in Fig. \ref{fig : slow_regime}. Substituting this value into 
Eq. \eqref{eq:scaling_slow} yields an estimate of $\beta \approx 0.015$.

We now shift our focus to the fast erosion regime. Our objective is to 
determine the functional form of $\mathcal{F}$, which allows for the calculation 
of the surface erosion rate $K_e$ as a function of the dimensionless 
group $I/\xi$ and the characteristic timescale $\tau_f = I \sqrt{\xi}/\dot{\gamma}$.

\subsection{Analysis in the fast erosion regime} 
\label{sec:sim fast}

Figure \ref{fig : time evolution} indicates that the number of eroded particles 
follows a nonlinear evolution over time, with a rate that increases 
sharply with the inclination angle. This nonlinear progression of $N_e(t)$ is 
primarily attributed to the evolving height $h$ of the non-cohesive 
layer. As erosion proceeds, the increasing thickness of the dry layer 
modifies both the confining pressure $\sigma_p$ and the local shear 
rate $\dot{\gamma}$ at the interface, thereby creating a feedback loop in the erosion dynamics.

The evolution of the normalized number of eroded particles, $N_e(t)/N_s$,
 is detailed in Fig. \ref{fig : time evolution}(a) for a surface tension 
 $\gamma_s = 0.01$~mN/m at three distinct inclination angles 
 ($\theta = 25^\circ, 28^\circ$, and $32^\circ$). Conversely, Fig. \ref{fig : time evolution}(b) 
 presents the evolution of $N_e(t)/N_s$ at a fixed inclination of $\theta = 32^\circ$ 
 for surface tension values ranging from $\gamma_s = 8 \times 10^{-3}$~N/m to $4 \times 10^{-2}$~N/m. 
 Consistent with physical expectations, the erosion rate is enhanced by 
 both steeper inclination angles and reduced surface tension.

For each combination of surface tension $\gamma_s$ and inclination 
angle $\theta$, and at every time step during the erosion process, 
we extract the surface erosion rate $K_e$ from the data in Fig. \ref{fig : time evolution}. 
The instantaneous height $h$ and the corresponding shear rate 
$\dot{\gamma}$ at the wet-dry interface are simultaneously determined 
from the velocity profiles as those shown in Fig. \ref{fig:VelocityProfiles}.

\begin{figure}[!tbh]
\begin{center}
\includegraphics[width=0.90\columnwidth]{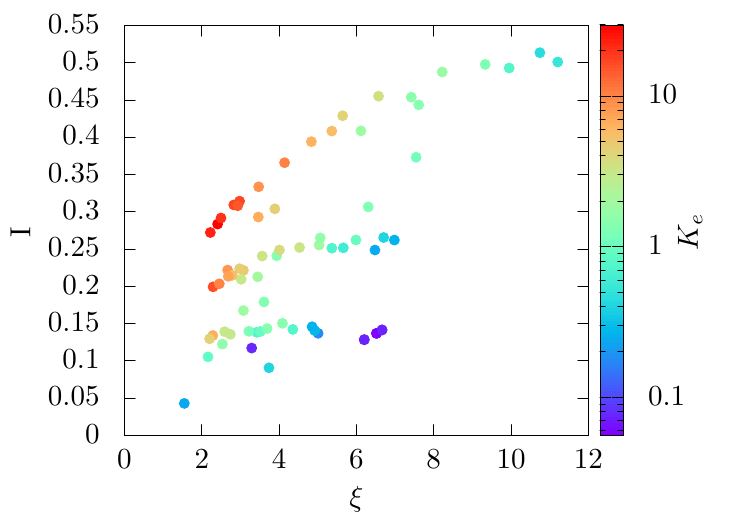}
\end{center}
\caption{Dimensionless mapping of the surface erosion rate $K_e = \frac{1}{N_s} \frac{dN_e}{dt}$ 
as a function of the interfacial inertial number $I$ and the cohesion index $\xi$. 
The data set comprises results from simulations spanning surface 
tensions $\gamma_s$ from $8 \times 10^{-3}$~N/m to $4 \times 10^{-2}$~N/m 
and inclination angles $\theta$ between $25^\circ$ and $32^\circ$. 
The color scale represents the magnitude of the erosion rate, 
illustrating the transition between different erosion intensities within the $I$-$\xi$ parameter space.}
\label{fig:Ke palette}
\end{figure}

Fig. \ref{fig:Ke palette} presents a comprehensive mapping of the 
inertial number $I$, the cohesion index $\xi$, and the resulting surface erosion 
rate $K_e$ (indicated by the color scale) for the entire set of simulations. 
The data points aggregated in this plot encompass the full range of 
inclination angles $\theta$ and surface tension values $\gamma_s$ investigated, 
captured at successive temporal intervals throughout each simulation. 

\begin{figure}[tbh]
\begin{center}
\includegraphics[width=0.90\columnwidth]{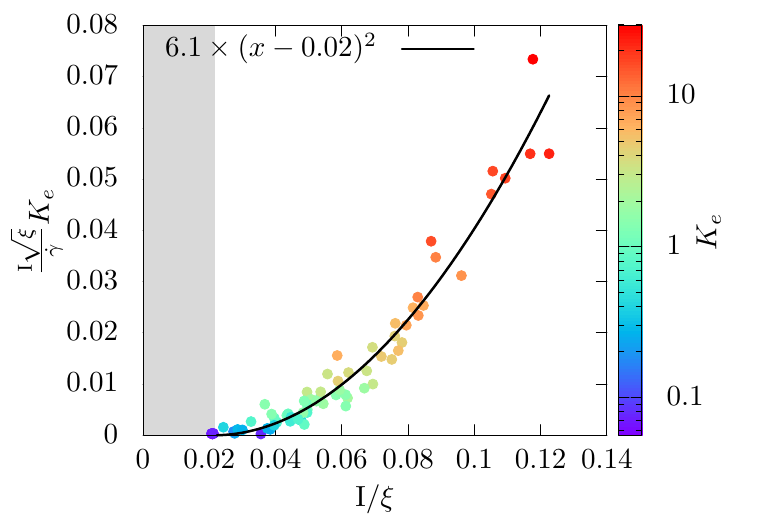}
\end{center}
\caption{Reduced erosion rate  $\frac{I\sqrt{\xi}}{\dot{\gamma}} K_e$ as 
a function of $I/\xi$. The solid curve is the result of a fit of the form $\frac{I\sqrt{\xi}}{\dot{\gamma}} 
K_e= A\left[(I/\xi)-(I/\xi)^\star\right]^\kappa$, with
$\kappa=1,97$,  $A=6$ and $(I/\xi)^\star=0.02$.
} 
\label{fig : fit}
\end{figure}

Following the scaling arguments developed in Eq. \eqref{eqn:scaling fast}, 
we examine the relationship between the surface erosion rate and the 
dimensionless control parameters. By plotting the product of the erosion 
rate and the characteristic timescale, $\tau_f K_e = \frac{I\sqrt{\xi}}{\dot{\gamma}} K_e$, 
as a function of the ratio $I/\xi$, we observe that all data points nicely collapse onto 
a single master curve, as shown in Fig. \ref{fig : fit}. This collapse, which holds 
across the entire range of investigated inclination angles and surface 
tension values in the fast erosion regime, confirms that $I/\xi$ is the fundamental 
dimensionless group governing the transition to and the intensity of the fast erosion regime.

The master curve is accurately described by a power-law relationship of the form:
\begin{equation}
\frac{I\sqrt{\xi}}{\dot{\gamma}} K_e = A \left[ \frac{I}{\xi} - \left( \frac{I}{\xi} \right)^\star \right]^\kappa, 
\label{eq:master_curve}
\end{equation}
where the best-fit parameters are found to be $\kappa \simeq 1.97$, 
$A \simeq 6$, and a critical threshold $(I/\xi)^\star \simeq 0.02$. 

The near-quadratic exponent ($\kappa \approx 2$) suggests a 
significant non-linear sensitivity of the erosion rate to the excess stress ratio.

To evaluate the consistency of our results, we compare the threshold 
value $(I/\xi)^\star$ obtained from the numerical fit with the theoretical 
prediction $a/\alpha$ derived from Eq. \eqref{eqn:Ixi}. Following the 
methodology of Vo et al. \cite{vo2020additive}, we adopt $a = 0.095$. 
The parameter $\alpha$ is estimated using the standard rheological values 
for hard spheres with an inter-particle friction coefficient $\mu_p = 0.5$ \cite{Jop2006constitutive}. 
Taking the values $I_0 \approx 0.3$, $\mu_0 \approx 0.4$, and $\mu_1 \approx 1.0$, 
we calculate $\alpha = (\mu_1 - \mu_0)/(\mu_0 I_0) \approx 5$. This leads to 
a theoretical ratio $a/\alpha \approx 0.019$. This value is in agreement with 
the threshold $(I/\xi)^\star \simeq 0.02$ determined from our data collapse.

It should be noted that certain physical parameters, such as the particle 
diameter $d$, density $\rho$, and gravitational acceleration $g$, were 
held constant throughout this study. Furthermore, the established relationships 
were obtained for fixed values of the capillary bridge rupture distance 
and the wetting angle. A complementary investigation exploring the influence 
of these additional parameters would be of significant value to further generalize these findings.

By substituting the definitions of the inertial number and cohesion index, 
the threshold for the fast erosion regime can be reformulated directly 
in terms of the mechanical stresses. Specifically, the condition for the onset of fast erosion becomes:
\begin{equation}
    \sqrt{\sigma_k \sigma_p} \geq b \sigma_c,
    \label{eqn:geometric}
\end{equation}
where $b = (I/\xi)^\star \simeq 0.02$. This expression highlights the fundamental 
interplay between the destabilizing stresses—the kinetic stress $\sigma_k$ and the 
confining pressure $\sigma_p$—and the resisting cohesive strength $\sigma_c$. 
Equation \eqref{eqn:geometric} reveals that the erosion threshold is governed 
by the ratio of the geometric mean of the driving stresses to the cohesive resistance. 
Remarkably, this suggests that the kinetic and confining stresses contribute with 
equal weight to the destabilization of the interface. 
This criterion can also be viewed as a limit on the sustainable cohesion for a given flow state:
\begin{equation}
    \sigma_c \leq \frac{I \sigma_p}{b}.
\end{equation}
Furthermore, by introducing the effective friction coefficient 
$\mu(I) = \sigma_\tau / \sigma_p$, where $\sigma_\tau$ is the shear stress at the interface, 
the erosion condition can be expressed as a function of the tangential loading:
\begin{equation}
    \sigma_\tau \geq \frac{b \mu(I) \sigma_c}{I}.
    \label{eqn:tangential}
\end{equation}
This formulation underscores the role of shear stress in the erosion process. 
More importantly, Eq. \eqref{eqn:tangential} demonstrates that a simple 
comparison between shear stress and cohesive strength is insufficient to 
characterize the erosion threshold; the dynamical state of the flow, 
encapsulated by the inertial number $I$, must also be explicitly considered 
to account for the mechanical stability of the interface.

\section{Conclusion} 
\label{sec:conclusion}

In this study, we have employed DEM simulations to investigate the 
microscopic mechanisms governing the erosion of a cohesive granular 
bed subjected to the shear flow of a non-cohesive dry layer. 
By modeling inter-particle cohesion through capillary bridge forces, 
we characterized erosion as the irreversible rupture of these liquid bonds 
at the interface. Our results reveal that the transition from stability to entrainment 
is not merely a function of a single stress threshold, but involves rather a complex 
interplay between the dynamical state of the dry flow and the mechanical resistance of the wet bed.

A key contribution of this work is the identification of two distinct 
erosion regimes: the first regime is a slow, stochastic regime where the 
mean driving stresses are below the yield threshold. In this case, we hypothesized 
that erosion is  activated by high-energy fluctuations in the granular 
temperature of the dry flow, following an Arrhenius-like probability law. 
The second regime is a fast, collective regime characterized by a mean-field 
mechanical instability of the interface.

Through systematic variations of the inclination angle and surface tension, 
we demonstrated that the surface erosion rate $K_e$ in the fast erosion 
regime is governed by a robust scaling law. By introducing a characteristic 
timescale $\tau_f$ derived from the competition between confining pressure 
and cohesive energy, we achieved a remarkable data collapse. 
This collapse highlights that the ratio of the inertial number $I$ to the 
cohesion index $\xi$ is the fundamental dimensionless group controlling 
the process. Our findings, summarized by the condition $\sqrt{\sigma_k \sigma_p} \geq b \sigma_c$, 
prove that the geometric mean of the kinetic and confining stresses must 
overcome the cohesive strength to initiate rapid erosion. 
This result underscores that shear stress alone is insufficient to characterize 
the erosion threshold; the dynamical inertia of the flow must be explicitly accounted for.

While our model provides a clear scaling framework by neglecting 
liquid viscosity, it serves as a baseline for more complex fluid-grain interactions. 
The scaling laws established here offer a predictive tool for broader 
contexts where cohesive media are eroded by granular flows, including 
industrial transport and geophysical events where driving forces may stem 
from sources other than gravity.

Future research should aim to extend this framework by incorporating 
lubrication forces and viscous effects within the capillary bridges. 
Such additions would be essential for capturing regimes where high shear 
rates or high liquid viscosities make viscous dissipation comparable 
to capillary attraction. Furthermore, exploring the reversibility of capillary 
bonds—allowing for the re-agglomeration of eroded particles—could provide 
a more nuanced understanding of mass balance in closed granular systems.


\begin{thebibliography}{10}

\bibitem{herminghaus2005dynamics}
Stephan Herminghaus.
\newblock Dynamics of wet granular matter.
\newblock {\em Advances in physics}, 54(3):221--261, 2005.

\bibitem{mitarai2006wet}
Namiko Mitarai and Franco Nori.
\newblock Wet granular materials.
\newblock {\em Advances in physics}, 55(1-2):1--45, 2006.

\bibitem{vo2020evolution}
Thanh-Trung Vo, Patrick Mutabaruka, Saeid Nezamabadi, Jean-Yves Delenne, and
  Farhang Radjai.
\newblock Evolution of wet agglomerates inside inertial shear flow of dry
  granular materials.
\newblock {\em Physical Review E}, 101(3):032906, 2020.

\bibitem{Lefebvre2013Erosion}
Gautier Lefebvre and Pierre Jop.
\newblock Erosion dynamics of a wet granular medium.
\newblock {\em Phys. Rev. E}, 88:032205, Sep 2013.

\bibitem{radjai2010force}
Farhang Radjai, Vincent Topin, Vincent Richefeu, Charles Voivret, Jean-Yves
  Delenne, Emilien Az{\'e}ma, and Said El~Youssoufi.
\newblock Force transmission in cohesive granular media.
\newblock In {\em AIP Conference Proceedings}, volume 1227, pages 240--259.
  American Institute of Physics, 2010.

\bibitem{bagnold1954experiments}
Ralph~Alger Bagnold.
\newblock Experiments on a gravity-free dispersion of large solid spheres in a
  newtonian fluid under shear.
\newblock {\em Proceedings of the Royal Society of London. Series A.
  Mathematical and Physical Sciences}, 225(1160):49--63, 1954.

\bibitem{da2005rheophysics}
Fr{\'e}d{\'e}ric Da~Cruz, Sacha Emam, Micha{\"e}l Prochnow, Jean-No{\"e}l Roux,
  and Fran{\c{c}}ois Chevoir.
\newblock Rheophysics of dense granular materials: Discrete simulation of plane
  shear flows.
\newblock {\em Physical Review E}, 72(2):021309, 2005.

\bibitem{iveson2002dynamic}
Simon~M Iveson, Jai~A Beathe, and Neil~W Page.
\newblock The dynamic strength of partially saturated powder compacts: the
  effect of liquid properties.
\newblock {\em Powder Technology}, 127(2):149--161, 2002.

\bibitem{moller2007shear}
Peder~CF M{\o}ller and D~Bonn.
\newblock The shear modulus of wet granular matter.
\newblock {\em Europhysics Letters}, 80(3):38002, 2007.

\bibitem{rognon2006rheophysics}
PG~Rognon, J-N Roux, D~Wolf, M~Naa{\"\i}m, and F~Chevoir.
\newblock Rheophysics of cohesive granular materials.
\newblock {\em Europhysics Letters}, 74(4):644, 2006.

\bibitem{richefeu2006shear}
Vincent Richefeu, Moulay~Sa{\"\i}d El~Youssoufi, and Farhang Radjai.
\newblock Shear strength properties of wet granular materials.
\newblock {\em Physical Review E}, 73(5):051304, 2006.

\bibitem{Braysh2025JOR}
Lama Braysh, Patrick Mutabaruka, Farhang Radjai, and Serge Mora.
\newblock Dynamics of wet granular flows down an inclined plane.
\newblock {\em Journal of Rheology}, 69:409--422, 2025.

\bibitem{GDR-MiDi2004}
GDR-MiDi.
\newblock On dense granular flows.
\newblock {\em Eur. Phys. J. E}, 14:341--365, 2004.

\bibitem{Jop2006constitutive}
Pierre Jop, Yo{\"e}l Forterre, and Olivier Pouliquen.
\newblock A constitutive law for dense granular flows.
\newblock {\em Nature}, 441(7094):727--730, Jun 2006.

\bibitem{kamrin2012nonlocal}
Ken Kamrin and Georg Koval.
\newblock Nonlocal constitutive relation for steady granular flow.
\newblock {\em Physical review letters}, 108(17):178301, 2012.

\bibitem{berger2016scaling}
Nicolas Berger, Emilien Az{\'e}ma, Jean-Fran{\c{c}}ois Douce, and Farhang
  Radjai.
\newblock Scaling behaviour of cohesive granular flows.
\newblock {\em Europhysics Letters}, 112(6):64004, 2016.

\bibitem{khamseh2015flow}
Saeed Khamseh, Jean-No{\"e}l Roux, and Fran{\c{c}}ois Chevoir.
\newblock Flow of wet granular materials: A numerical study.
\newblock {\em Physical Review E}, 92(2):022201, 2015.

\bibitem{badetti2018rheology}
M~Badetti, A~Fall, D~Hautemayou, F~Chevoir, Patrick Aimedieu, S~Rodts, and J-N
  Roux.
\newblock Rheology and microstructure of unsaturated wet granular materials:
  Experiments and simulations.
\newblock {\em Journal of rheology}, 62(5):1175--1186, 2018.

\bibitem{pouliquen2006flow}
Olivier Pouliquen, Cyril Cassar, Pierre Jop, Yoel Forterre, and Maxime Nicolas.
\newblock Flow of dense granular material: towards simple constitutive laws.
\newblock {\em Journal of Statistical Mechanics: Theory and Experiment},
  2006(07):P07020, 2006.

\bibitem{forterre2008flows}
Yo{\"e}l Forterre and Olivier Pouliquen.
\newblock Flows of dense granular media.
\newblock {\em Annu. Rev. Fluid Mech.}, 40:1--24, 2008.

\bibitem{badetti2018shear}
Michel Badetti, Abdoulaye Fall, Fran{\c{c}}ois Chevoir, and Jean-No{\"e}l Roux.
\newblock Shear strength of wet granular materials: Macroscopic cohesion and
  effective stress: Discrete numerical simulations, confronted to experimental
  measurements.
\newblock {\em The European Physical Journal E}, 41:1--16, 2018.

\bibitem{boyer2011unifying}
Fran{\c{c}}ois Boyer, {\'E}lisabeth Guazzelli, and Olivier Pouliquen.
\newblock Unifying suspension and granular rheology.
\newblock {\em Physical review letters}, 107(18):188301, 2011.

\bibitem{trulsson2012transition}
Martin Trulsson, Bruno Andreotti, and Philippe Claudin.
\newblock Transition from the viscous to inertial regime in dense suspensions.
\newblock {\em Physical review letters}, 109(11):118305, 2012.

\bibitem{vo2020additive}
Thanh~Trung Vo, Saeid Nezamabadi, Patrick Mutabaruka, Jean-Yves Delenne, and
  Farhang Radjai.
\newblock Additive rheology of complex granular flows.
\newblock {\em Nature communications}, 11(1):1476, 2020.

\bibitem{vo2021erosion}
Thanh-Trung Vo.
\newblock Erosion dynamics of wet particle agglomerates.
\newblock {\em Computational Particle Mechanics}, 8(3):601--612, 2021.

\bibitem{amarsid2017viscoinertial}
Lhassan Amarsid, J-Y Delenne, Patrick Mutabaruka, Yann Monerie,
  Fr{\'e}d{\'e}ric Perales, and Farhang Radjai.
\newblock Viscoinertial regime of immersed granular flows.
\newblock {\em Physical Review E}, 96(1):012901, 2017.

\bibitem{Tapia2022}
Franco Tapia, Mie Ichihara, Olivier Pouliquen, and \'Elisabeth Guazzelli.
\newblock Viscous to inertial transition in dense granular suspension.
\newblock {\em Phys. Rev. Lett.}, 129:078001, Aug 2022.

\bibitem{hassanpour2007effect}
A~Hassanpour, SJ~Antony, and M~Ghadiri.
\newblock Effect of size ratio on the behaviour of agglomerates embedded in a
  bed of particles subjected to shearing: Dem analysis.
\newblock {\em Chemical engineering science}, 62(4):935--942, 2007.

\bibitem{hassanpour2007modeling}
Ali Hassanpour, S~Joseph Antony, and Mojtaba Ghadiri.
\newblock Modeling of agglomerate behavior under shear deformation: effect of
  velocity field of a high shear mixer granulator on the structure of
  agglomerates.
\newblock {\em Advanced Powder Technology}, 18(6):803--811, 2007.

\bibitem{mangeney2010erosion}
A~Mangeney, Olivier Roche, O~Hungr, N~Mangold, Gloria Faccanoni, and A~Lucas.
\newblock Erosion and mobility in granular collapse over sloping beds.
\newblock {\em Journal of Geophysical Research: Earth Surface}, 115(F3), 2010.

\bibitem{He2024}
Xurong He, Xiewen Hu, Zihao Huo, Jianfeng Tang, and Shilin Zhang.
\newblock Study on dynamic mechanism of granular flow erosion and entrainment
  based on dem theory.
\newblock {\em Geoenviron Disasters}, 11:1--13, {article no.} 17, 2024.

\bibitem{lajeunesse2010bed}
Eric Lajeunesse, Luce Malverti, and Fran{\c{c}}ois Charru.
\newblock Bed load transport in turbulent flow at the grain scale: Experiments
  and modeling.
\newblock {\em Journal of Geophysical Research: Earth Surface}, 115(F4), 2010.

\bibitem{charru2004erosion}
Fran{\c{c}}ois Charru, H{\'e}lene Mouilleron, and Olivier Eiff.
\newblock Erosion and deposition of particles on a bed sheared by a viscous
  flow.
\newblock {\em Journal of Fluid Mechanics}, 519:55--80, 2004.

\bibitem{ennis1991microlevel}
Bryan~J Ennis, Gabriel Tardos, and Robert Pfeffer.
\newblock A microlevel-based characterization of granulation phenomena.
\newblock {\em Powder Technology}, 65(1-3):257--272, 1991.

\bibitem{behjani2017investigation}
Mohammadreza~Alizadeh Behjani, Nejat Rahmanian, Ali Hassanpour, et~al.
\newblock An investigation on process of seeded granulation in a continuous
  drum granulator using dem.
\newblock {\em Advanced Powder Technology}, 28(10):2456--2464, 2017.

\bibitem{Mutabaruka2014Initiation}
Patrick Mutabaruka, Jean-Yves Delenne, Kenichi Soga, and Farhang Radjai.
\newblock Initiation of immersed granular avalanches.
\newblock {\em Phys. Rev. E}, 89:052203, May 2014.

\bibitem{Mutabaruka2019Effects}
Patrick Mutabaruka, Mahdi Taiebat, Roland J.-M. Pellenq, and Farhang Radjai.
\newblock Effects of size polydispersity on random close-packed configurations
  of spherical particles.
\newblock {\em Phys. Rev. E}, 100:042906, Oct 2019.

\bibitem{cundall1979discrete}
Peter~A Cundall and Otto~DL Strack.
\newblock A discrete numerical model for granular assemblies.
\newblock {\em geotechnique}, 29(1):47--65, 1979.

\bibitem{herrmann1998modeling}
HJ~Herrmann and Stefan Luding.
\newblock Modeling granular media on the computer.
\newblock {\em Continuum Mechanics and Thermodynamics}, 10:189--231, 1998.

\bibitem{thornton2000quasi}
C~Thornton and SJ~Antony.
\newblock Quasi-static shear deformation of a soft particle system.
\newblock {\em Powder technology}, 109(1-3):179--191, 2000.

\bibitem{radjai2011discrete}
Farhang Radja\"\i and Fr\'ed\'eric Dubois, editors.
\newblock {\em Discrete-element modeling of granular materials}.
\newblock Iste-Wiley, New-York, March 2011.
\newblock ISBN: 978-1-84821-260-2.

\bibitem{allen1987computer}
Michael~P Allen, Dominic~J Tildesley, et~al.
\newblock Computer simulation of liquids.
\newblock {\em Clarendon-0.12}, 1987.

\bibitem{moreau1994sorne}
Jean~Jacques Moreau.
\newblock Sorne numerical methods in multibody dynamics: application to
  granular materials.
\newblock {\em European Journal of Mechanics-A/Solids}, 13(4-suppl):93--114,
  1994.

\bibitem{radjai2009contact}
Farhang Radjai and Vincent Richefeu.
\newblock Contact dynamics as a nonsmooth discrete element method.
\newblock {\em Mechanics of Materials}, 41(6):715--728, 2009.

\bibitem{radjai2018multi}
Farhang Radjai.
\newblock Multi-periodic boundary conditions and the contact dynamics method.
\newblock {\em Comptes Rendus M{\'e}canique}, 346(3):263--277, 2018.

\bibitem{herrmann2013physics}
Hans~J Herrmann, J-P Hovi, and S~Luding.
\newblock {\em Physics of dry granular media}, volume 350.
\newblock Springer Science \& Business Media, 2013.

\bibitem{scheel2008liquid}
Mario Scheel, Ralf Seemann, Martin Brinkmann, Marco Di~Michiel, Adrian
  Sheppard, and Stephan Herminghaus.
\newblock Liquid distribution and cohesion in wet granular assemblies beyond
  the capillary bridge regime.
\newblock {\em Journal of Physics: Condensed Matter}, 20(49):494236, 2008.

\bibitem{Delenne2015}
Jean-Yves Delenne, Vincent Richefeu, and Farhang Radjai.
\newblock Liquid clustering and capillary pressure in granular media.
\newblock {\em Journal of Fluid Mechanics}, 762(R5):R5, 2015.

\bibitem{Braysh2024JFM}
Lama Braysh, Patrick Mutabaruka, Farhang Radjai, and Serge Mora.
\newblock Breakage dynamics and scaling of wet aggregates of rigid particles.
\newblock {\em Journal of Fluid Mechanics}, 1000:A39, 2024.

\bibitem{Mikami1998}
T.~Mikami, H.~Kamiya, and M.~Horio.
\newblock Numerical simulation of cohesive powder behavior in fluidized bed.
\newblock {\em Chemical Engineering Science}, 53(10):1927--1940, 1998.

\end{thebibliography}

\end{document}